\documentclass[5p]{elsarticle}

\usepackage{lineno,hyperref}
\usepackage{siunitx}
\usepackage{hyperref}
\usepackage{mathtools}
\usepackage{rotating}
\usepackage{multirow}
\usepackage{xcolor,colortbl}
\usepackage{csquotes}
\usepackage{soul}

\modulolinenumbers[5]

\journal{Nuclear Instruments and Methods in Physics Research Section A, }

\begin{document}

\begin{frontmatter}

\title{An automated testing system for the RD51 VMM hybrid and yield measurement of the first production batches}


\author[a]{Finn Jaekel\corref{TUD}}
\cortext[TUD]{Now with 
Technische Universit\"at Dresden}
\author[a]{Klaus Desch}
\author[a]{Jochen Kaminski}
\author[a]{Michael Lupberger\corref{mycorrespondingauthor}}
\cortext[mycorrespondingauthor]{Corresponding author}\ead{lupberger@physik.uni-bonn.de}
\author[a,b]{Lucian Scharenberg}
\author[a]{Patrick Schwaebig}

\address[a]{Universit\"at Bonn, Bonn, Germany}
\address[b]{CERN, Geneva, Switzerland}

\begin{abstract}
We present the development of an automated testing system for the VMM hybrid of the RD51 collaboration. The VMM hybrid is a new front-end board for the RD51 common readout system, the Scalable Readout System, and will become the workhorse for the next decade to read out Micro-Pattern Gaseous Detectors. It uses the VMM chip developed for the ATLAS New Small Wheel to convert charge signals from detectors to digital data. Our testing system automatically characterises the quality of the VMM chips on the hybrid after production during multiple tests. Results are evaluated to classify the chips and hybrids and uploaded to a database. We evaluated those results for the first two production batches to measure the production yield. The yield is better than the threshold below which chip testing on a wafer level offers financial benefits. Observations on prominent chip failures were propagated back to the hybrid production process to further increase the yield for future productions.
\end{abstract}


\end{frontmatter}


\section{Introduction}\label{sec:Intro}
The Scalable Readout System (SRS)~\cite{SRS3} is the general-purpose readout electronics developed by the RD51 collaboration. It implements several front-end Application-Specific Integrated Circuits (ASICs) to read out mainly gaseous detectors. The scalability is achieved in a star-like architecture. The computer system with data acquisition and slow control software can be connected to several Front-End Concentrator (FEC) cards via Gigabit Ethernet. Those cards are the same for any front-end ASIC and can be interconnected to an ASIC specific adapter card by PCI express connectors. The adapter card connects several front-end boards by HDMI cables and the communication is based on a specific data protocol.

The system was used in a wide range of applications e.g.~\cite{APV1, APV2, APV3, APV4} mainly with the APV25~\cite{APV25} as front-end ASIC. To cope with the requirements of the latest experiments as e.g. high rate and triggerless readout, a new front-end ASIC, the VMM~\cite{VMM1} was implemented in the SRS recently~\cite{SRS_VMM}. Within the next decade, the SRS with the VMM is expected to enable R\&D projects and the instrumentation of upcoming experiments. The system has left its basic development phase and the first production runs of VMM front-end cards, called VMM hybrids, have been completed. There is an increasing number of projects, within which the VMM hybrid of the RD51 collaboration will be applied. Besides R\&D projects mainly in the domain of gaseous detectors, the board will be used for detectors at the European Spallation Source~\cite{ESS3}, the MAGIX experiment~\cite{MAGIX} and geographic muon tomography~\cite{MuonTomo}, to list only a few.

Due to the increasing hybrid production volumes, which have reached quantities of about one thousand per year, testing by hand is no longer feasible. Two automated test systems have been developed. The VMM Testing System (VTC)~\cite{VTC} is indispensable during the production process and can detect severe defects like short circuits or faulty component soldering as well as major VMM ASICs issues. The other system, reported in this manuscript, is meant for a complete characterisation of the signal path from the connector to the detector up to the data output connector of the hybrid. It classifies in particular the VMM ASICs by a large variety of testing procedures and is used to determine the VMM hybrid production yield. By using the standard SRS readout chain, the hybrids and their VMM chips can be qualified and re-qualified at any time when being used for experimental studies.

The VMM ASIC and hybrid are introduced in section~\ref{sec:VMM}, followed by an introduction of the test platform in section~\ref{sec:SetupHardware}. Results of automated testing of the latest hybrid production batches, in particular the yields, are presented in section~\ref{sec:Results}, followed by the impact of the results and an outlook in section~\ref{sec:Impact}. The conclusion is provided in section~\ref{sec:Conc}.
\section{VMM and VMM hybrid}\label{sec:VMM}
In the scope of the ATLAS New Small Wheel (NSW) upgrade~\cite{NSW} a new ASIC, the VMM~\cite{VMM3a}, has been designed. It is used as a readout chip for the Micromegas~\cite{Giomataris} and small-strip Thin Gap Chamber (sTGC)~\cite{sTGC} detectors of the NSW, which was installed in the ATLAS experiment during the Long Shutdown 2 of the Large Hadron Collider at the European Organisation for Nuclear Research. The implementation in the SRS, however, aims at a general application. The VMM, designed in \SI{130}{\nano\meter} CMOS technology by Brookhaven National Laboratory, is an analogue to digital ASIC with 64 input channels. It digitises the charge signals from detectors and outputs arrival time, amplitude and channel number in a data stream of up to \SI{800}{Mbit/s}. In a single channel, hits with an arrival time difference of less than \SI{500}{\nano\second} can be separated. The analogue part of the VMM can be tuned by means of shaper peaking times of \SI{25}{}, \SI{50}{}, \SI{100}{} and \SI{200}{\nano\second}, preamplifier gains from \SI{0.5}{} to \SI{16}{\milli\volt\slash\femto\coulomb} and different time-to-digital precision. The chip is designed for a detector capacitance from a few pF up to 3 nF.

\begin{figure}[htbp]
    \centering
    \includegraphics[width=0.9\linewidth]{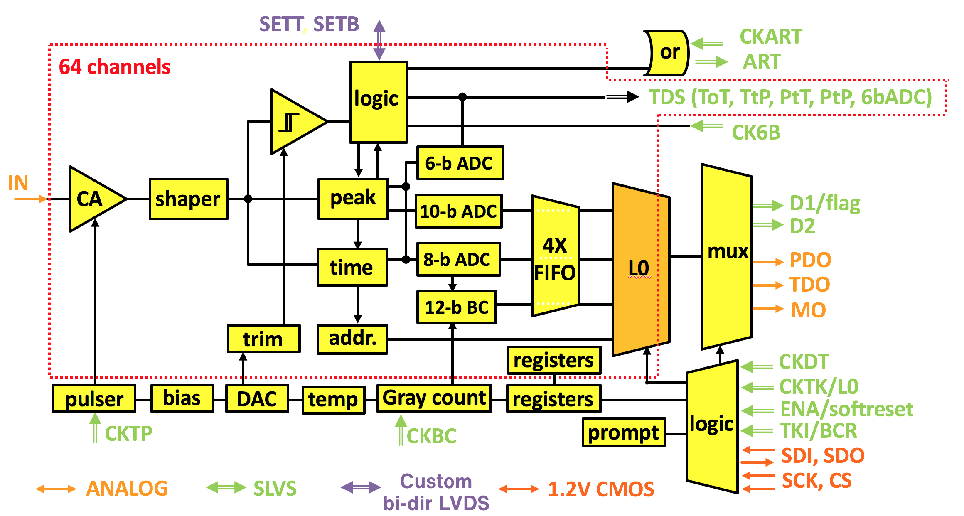}
    \caption{Block diagram of a single VMM ASIC \cite{VMM_manual}.}
    \label{fig:VMMBlockDiagram}
\end{figure}

After the preamplifier and shaper, the analogue signal is digitised if a configurable threshold is passed. This threshold is set globally for all channels. For each individual channel, the threshold can be fine adjusted by a five-bit trim Digital to Analogue Converter (DAC). For each threshold crossing, the digital part of the ASIC generates hit information. In order to not saturate the output bandwidth of the VMM, the threshold has to be set well above the baseline, such that only very few noise signals cross the threshold. The baseline level includes noise from the VMM internal circuits, sources on the Printed Circuit Board (PCB) it is mounted on as well as of the capacitance connected to the input. A block diagram of a VMM ASIC is shown in figure~\ref{fig:VMMBlockDiagram}. 

\begin{figure}[htbp]
    \centering
    \includegraphics[width=0.9\linewidth]{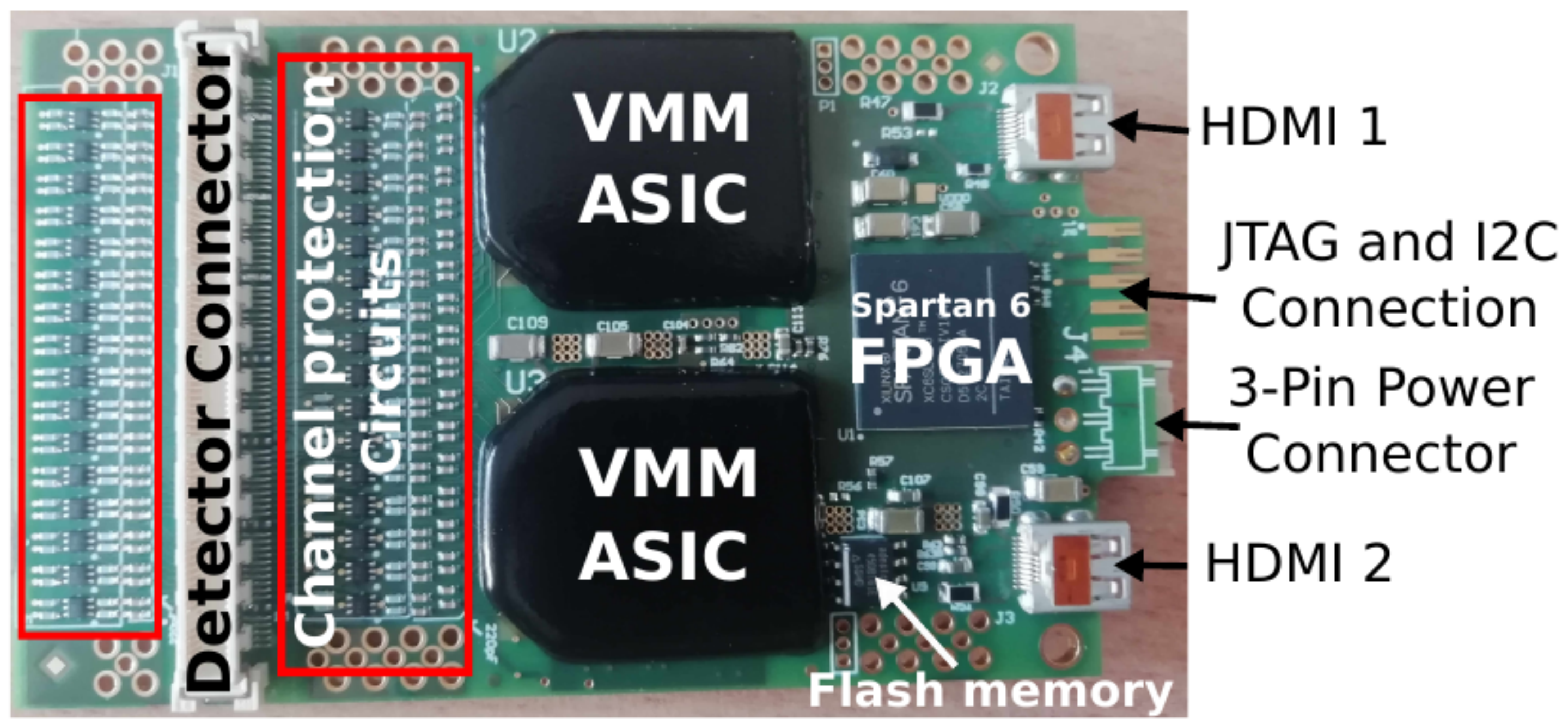}
    \caption{VMM hybrid, from \cite{Jaekel}}
    \label{fig:VMMHybrid}
\end{figure}

In the case of the VMM-based SRS readout, the front-end board, called hybrid, see figure~\ref{fig:VMMHybrid}, holds two VMM ASICs to be compatible and exchangeable with the 128 input channel APV25 SRS hybrid. Data of both VMMs are combined in a Xilinx Spartan-6 Field Programmable Gate Array (FPGA), which also controls the ASICs and other components on the board. In the future, the FPGA will be upgraded to a Spartan-7. The board can be either powered externally or through the HDMI cables used to connect the hybrid to the SRS back-end. Low Drop Out regulators (LDOs) on the backside of the hybrid assure a stable power supply. The voltage drop on those components adds about \SI{1}{\watt} to the \SI{2}{\watt} produced by the two VMM AISCs. For this reason, the hybrid comes with a cooling kit with convection radiators on the top side. An image of the VMM hybrid is shown in figure~\ref{fig:VMMHybrid}.

The hybrid also holds two Analogue to Digital Converters (ADCs) that can be read out by the I2C protocol. Each one connects to the Monitoring Output (MO), see figure~\ref{fig:VMMBlockDiagram} of one VMM chip and reads the analogue voltage level provided. The MO can be programmed to output several VMM internal voltage levels e.g. DAC value of the threshold, the pedestal for a selected channel or the voltage level of a VMM internal temperature sensor.

\section{Test platform setup and test procedure}\label{sec:SetupHardware}
In order to contribute to the production quality assurance and to provide a system to characterise VMM hybrids in a standardised, fast and automated way, a test platform was developed. The platform itself as well as the test procedure is explained in this chapter.
\subsection{Test platform hardware setup}\label{sec:Platform}
\begin{figure}[htbp]
    \centering
    \includegraphics[width=0.7\linewidth]{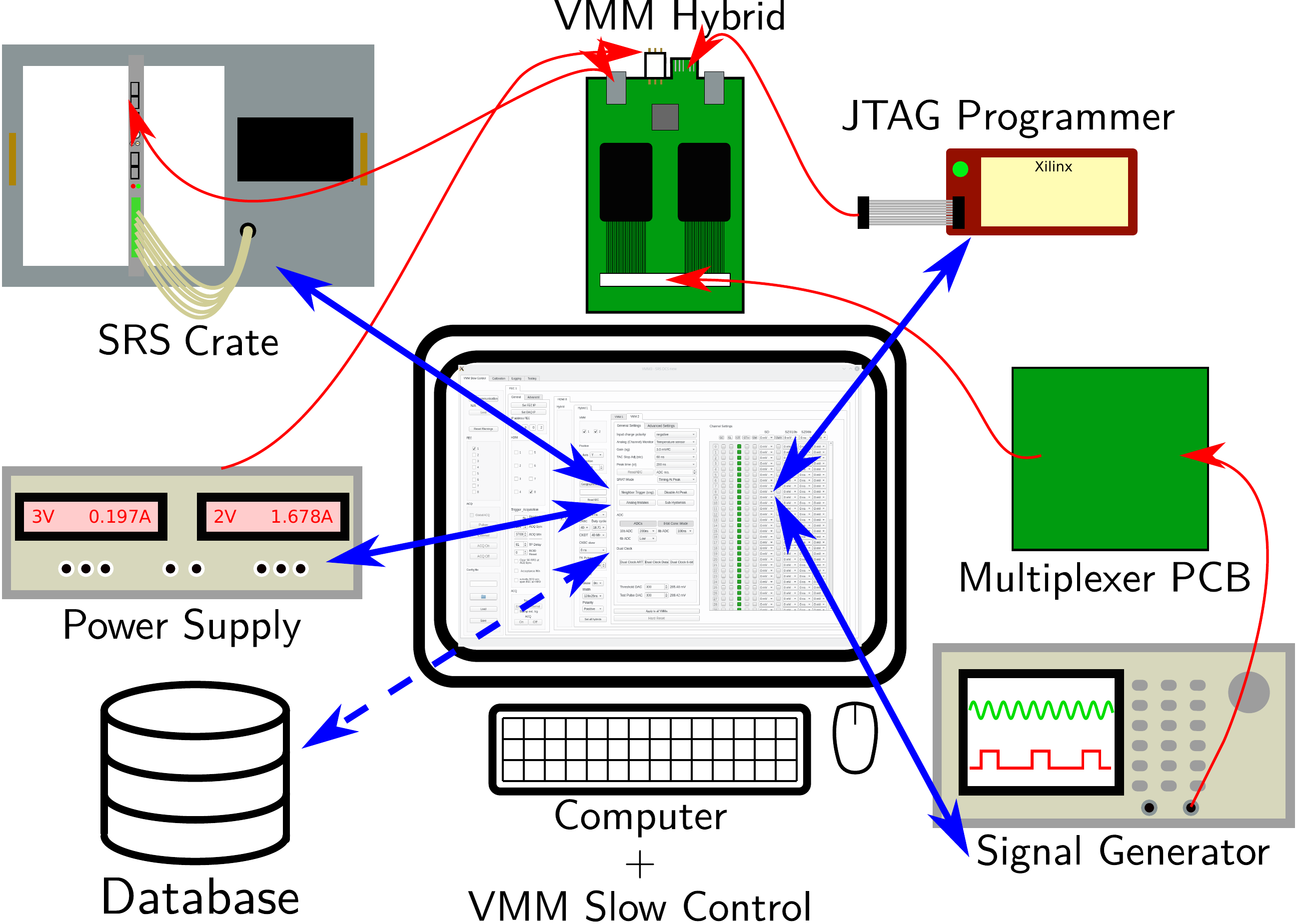}
    \caption{Schematic setup of a VMM hybrid testing station.}
    \label{fig:SchematicSetup}
\end{figure}

A schematic of the test setup is shown in figure \ref{fig:SchematicSetup}. The VMM hybrid under test is connected to the SRS, a JTAG programmer for firmware upload and optionally to a multiplexer PCB. The latter was specifically developed within this project. In combination with a remotely controllable signal generator, it allows the application of external test pulses to individual channels of the hybrid. Using an optional remote controllable power supply the supply current and voltages can be monitored. Powering through an HDMI connection directly from the FEC of the SRS is also possible.  All components are connected and controlled by a PC running the VMM slow control software \cite{VMMsc} (see section \ref{sec:SetupSoftware}). Data between the SRS and the PC are transferred via an Ethernet connection. Test results are automatically uploaded to a github repository, where all results are put into a single database file.

The power supply, the external test pulse system and the JTAG programmer are all optional components. A sufficient qualification and classification of the VMM hybrids can already be achieved using only a PC and the SRS with the proper adapter card and powering setup. 
\subsection{Test platform software}\label{sec:SetupSoftware}
\begin{figure*}[htbp]
    \centering
    \includegraphics[width=1.0\linewidth]{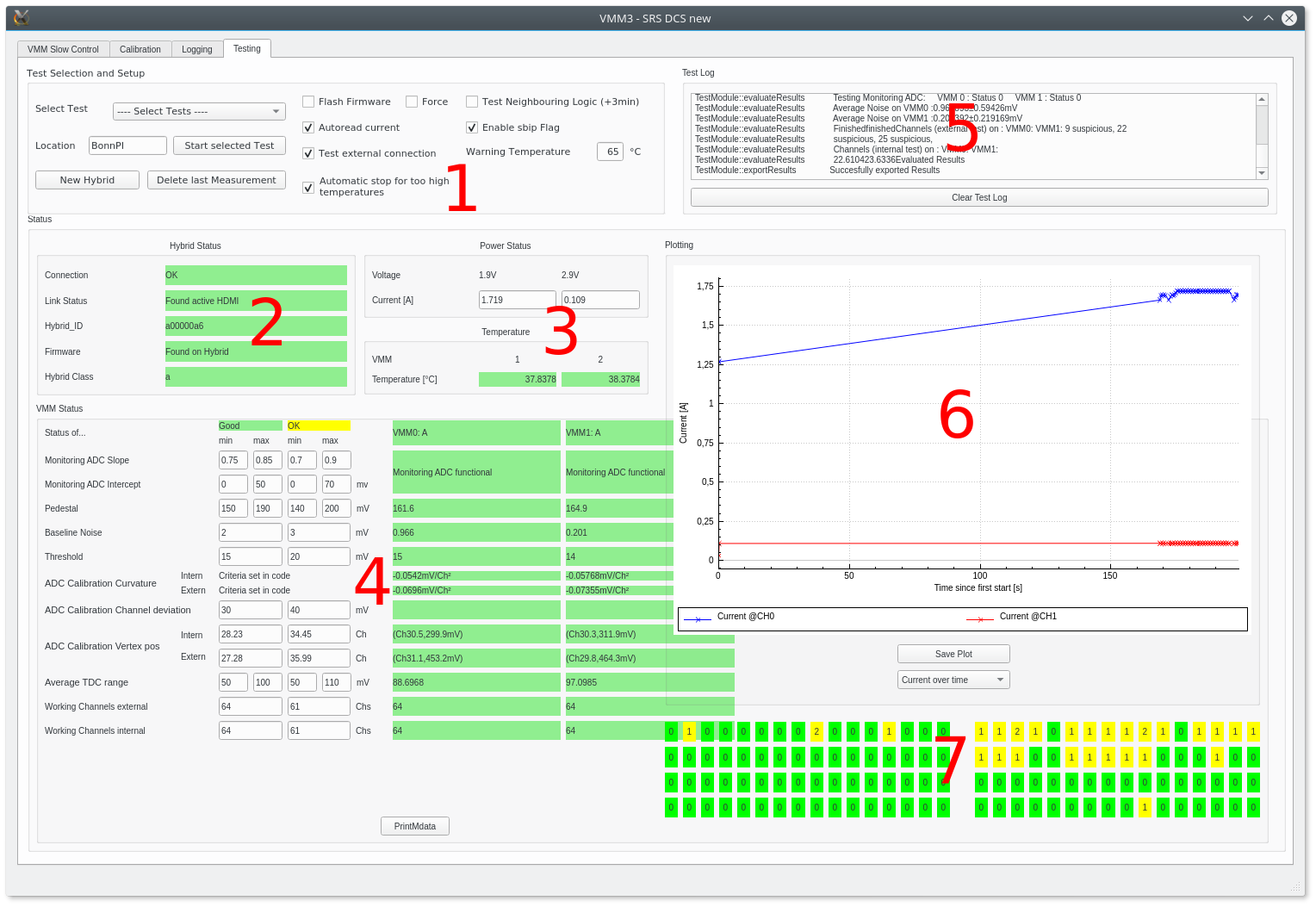}
    \caption{Graphical User Interface of the SRS slow control software tab for the VMM hybrid quality test, example of a hybrid test result. The numbers indicate the different parts: 1: Setup and test selection, 2: Hybrid status, 3: Temperature and power monitor, 4: VMM status, 5: Test log, 6: Plot window, 7: VMM channel status matrices.}
    \label{fig:GUI}
\end{figure*}

The VMM hybrid quality test was integrated by extending the VMM slow control software. Within the Graphical User Interface (GUI) shown in figure~\ref{fig:GUI}, all required settings and the automated test procedures can be selected. It also displays the test results. The SRS readout chain is used for collecting information and data about the hybrid. There are two different ways in which data can be gathered:
\begin{enumerate}
    \item Reading the ADC connected to the VMM MO via I2C (see section~\ref{sec:VMM})
    \item Evaluating digital VMM hit data as received on the PC (see section~\ref{sec:Platform})
\end{enumerate}

\subsection{Test procedure}
The hybrid to be characterised is connected to the FEC adapter card via HDMI, the external, remotely controlled power supply (optional), the multiplexer PCB (optional) and the JTAG programmer (optional). In case no external power supply is used, the hybrid is powered via the HDMI cable by the adapter card. Depending on the available setup, different options have to be selected in the slow control. E.g. in case an external, remotely controlled power supply is used, the power consumption test can be enabled (\textit{Autoread current} in figure~\ref{fig:GUI}). The physical location or institute the hybrids belongs to also needs to be entered there. All options can also be set in a configuration file which is automatically loaded at startup. The software automatically connects to the SRS and checks if an HDMI connection to a hybrid can be established. If no HDMI connection is found the software can, in case this option has been enabled and the JTAG programmer is connected, try to automatically load the hybrid firmware onto the FPGA and connect again. After reprogramming the FPGA, the programmer has to be disconnected, as for some VMM hybrids problems would occur otherwise. Afterwards, the hybrid ID\footnote{Only the least significant 64 bits of the 128-bit long hybrid ID, as these already uniquely classify a certain hybrid} is read from the EEPROM on the hybrid via the I2C protocol. A measurement ID is generated consisting of the hybrid ID and the \textit{datetime} in the format \texttt{yyyyMMddhhmm}.
Afterwards, the following tests are performed: 
\begin{enumerate}
\item Monitoring ADC Calibration: Increases the global VMM threshold DAC and reads out the monitoring output MO configured to the threshold via the monitoring ADC.
\item Pedestal and Pedestal Noise: Measures for each channel ten times the baseline and calculates the average, as well as the fluctuation of the baseline value.
\item Threshold: Measures the threshold of every channel
\item Threshold Trimability: Checks the functionality of the trimming threshold DAC for each individual channel
\item External Test Pulses: Evaluates the functionality of the channels using external test pulses
\item Internal Test Pulses: Evaluates the functionality of the channels using internal test pulses
\end{enumerate}
The first four tests in the list are using the monitoring ADC, the two test pulse tests use the SRS data acquisition chain. For both internal and external test pulses, there are subtests performed on the recorded data, such as the number of pulses recorded during the test, as well as the average ADC of all hits for each channel. Together with the pedestal, these are defined as critical tests.

After each test is completed, test specific quality criteria are evaluated for each channel\footnote{Except for the monitoring ADC Calibration, as this measurement is not channel but VMM specific.}. E.g. for the pedestal test, the average pedestal is checked to agree with the range of defined \texttt{good} and \texttt{ok} intervals. According to these criteria each channel is given an error score of either 0, 1 or 2 where 0 is given for \texttt{good} channels (pedestal value inside in the \texttt{good} interval), 1 for \texttt{ok} channels (pedestal value is outside the \texttt{good} interval, but within the \texttt{ok} interval) and 2 for \texttt{bad} channels (pedestal value is outside the \texttt{ok} interval). 

After all tests are finished, channels are classified as \texttt{broken} if they received an error score of 2 in one of the critical tests (see above). If the channels receive an error score of 2 or 1 in any other test, they are classified as \texttt{ok}, otherwise as \texttt{good}. 

It should be noted, that if the Monitoring ADC Calibration test fails for a VMM, the whole VMM is classified as bad, as in this case, no monitoring is possible. As during the production the hybrids are tested with the VTC, which filters out VMM with these problems, this problem should not occur.

Testing a single hybrid with all optional tests takes about two minutes. The data is automatically stored in the database.

\section{Results}\label{sec:Results}
\subsection{VMM and hybrid classification}
Depending on the total number of broken channels, the VMMs and hybrids are classified according to the classification as outlined in tables \ref{tab:ClassificationVMM} and \ref{tab:ClassificationHybrid}. Instead of a simple \enquote{working/broken} classification scheme, this slightly more detailed classification was chosen to allow to place hybrids of not perfect quality in less critical areas of a detector or use them for test purposes. The classification also accounts for the variety of problems which may occur on a hybrid.
\begin{table}[htbp]
    \centering
    \begin{tabular}{|c|c|}
 		\rowcolor[HTML]{000000} \color[HTML]{ffffff} VMM Class & \color[HTML]{ffffff}Number of broken Channels  \\\hline
		\rowcolor[HTML]{92d050} A & 0\\
		\rowcolor[HTML]{ff85ff} B & 1\\
		\rowcolor[HTML]{ffff00} C & 2-3 \\
		\rowcolor[HTML]{ffc000} D & 4-32\\
		\rowcolor[HTML]{ff0000} E & $>$32 
    \end{tabular}
    \caption{Classification table for the single VMM ASICs on a hybrid.}
    \label{tab:ClassificationVMM}
\end{table}
\begin{table}[htbp]
    \centering
	\begin{tabular}{|c|c|c|c|}\hline
		\rowcolor[HTML]{000000} \color[HTML]{ffffff} 
        \vtop{\hbox{\strut Hybrid}\hbox{\strut\hspace{0.5mm} class}} & \color[HTML]{ffffff} \vtop{\hbox{\strut VMM 1}\hbox{\strut \hspace{2.5mm}class}} & \color[HTML]{ffffff} \vtop{\hbox{\strut VMM 2}\hbox{\strut \hspace{2.5mm}class}} & \color[HTML]{ffffff} \vtop{\hbox{\strut Total Number of}\hbox{\strut broken Channels}} \\\hline
		\rowcolor[HTML]{92d050} a & A & A & 0\\
		\rowcolor[HTML]{ffff00} b & A/B & B/C/B & 1-3 \\
		\rowcolor[HTML]{ff0000} c & A/B/C/D & D/E &4-61\\
		\rowcolor[HTML]{ff0000} c & B/C & C &4-61\\
		\rowcolor[HTML]{000000} \color[HTML]{ffffff} d & \color[HTML]{ffffff} A/B/C/D/E & \color[HTML]{ffffff} E & \color[HTML]{ffffff} $>$61\\\hline
	\end{tabular}
    \caption{Classification for a VMM hybrid.}
    \label{tab:ClassificationHybrid}
\end{table}
\subsection{Pedestal problem}
\begin{figure}[htb]
    \centering
    \includegraphics[width=\linewidth]{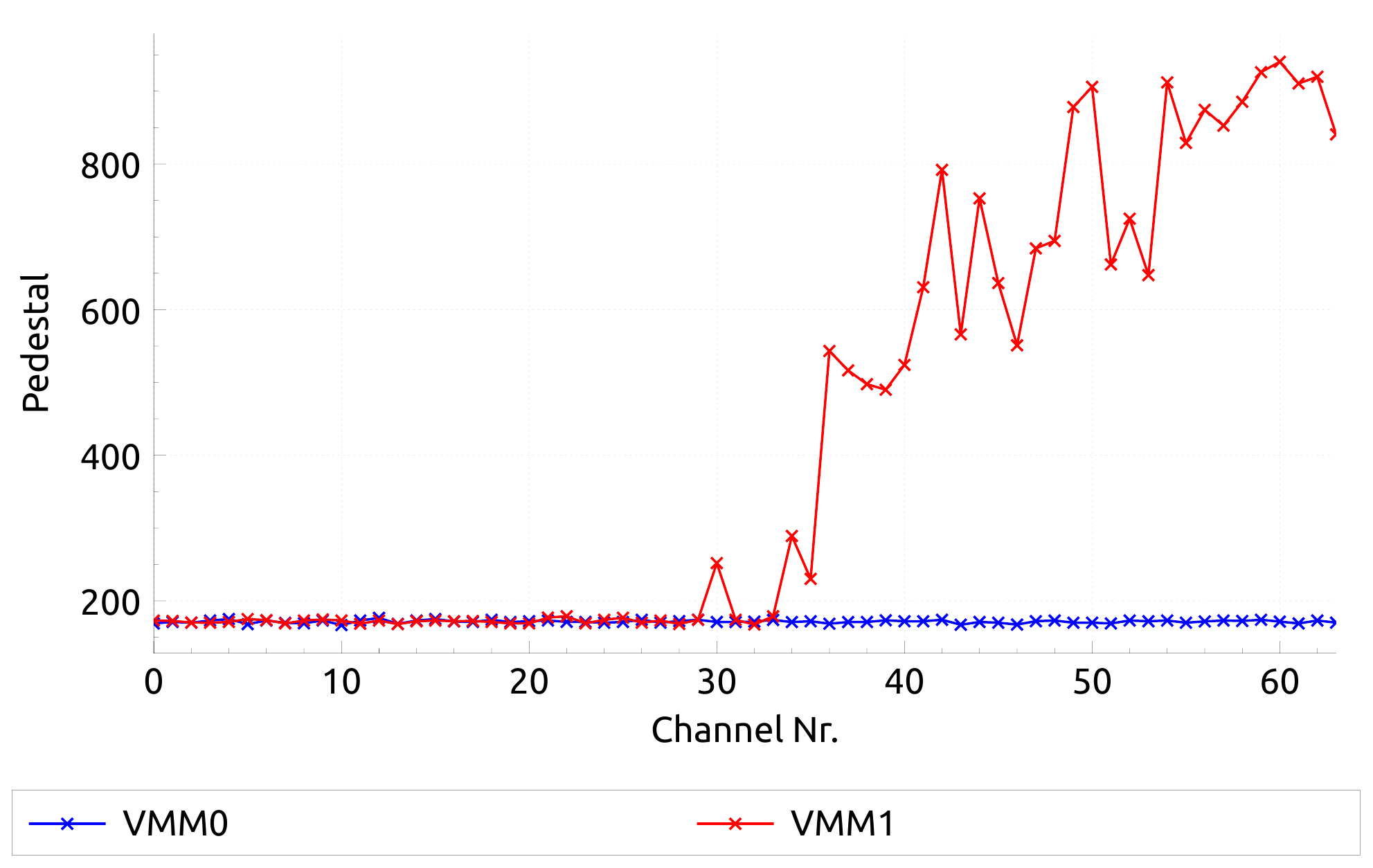}
    \caption{Example of for pedestal problem: Some VMM\,1 (red) channels have a high baseline level, when the highest gain is used. VMM\,0 shows a typical baseline at about \SI{160}{\milli\volt} for comparison.}
    \label{fig:pedestalProblem}
\end{figure}

A known error of the VMM3a can result in a malfunction of channels, which manifests in a high level of the channel baseline (pedestal)~\cite{VMM3a}. The baseline level settles down too slow after configuration and can last for a few \SI{}{\milli\second}. Identification is only possible if the pedestal is measured rapidly after changing the channel settings. The problem is more prominent at high preamplifier gains and can be circumvented by setting the configuration bits \textit{stlc} or \textit{sbip}, where the latter is recommended as it has a larger effect. The default configuration has \textit{sbip} enabled. However, it was observed during the testing that some VMM chips still have channels with high baselines at high gains down to \SI{6}{\milli\volt\per\femto\coulomb}. For a gain of \SI{3}{\milli\volt\per\femto\coulomb}, the baseline of all channels on all those chips showed a normal behaviour. Depending on the application's needs, those chips can be applied the same way as chips without this behaviour. However, a pedestal problem at high gains despite the enabling of \textit{sbip} could indicate a general chip issue. It was decided to introduce a dedicated marking for this behaviour.  A "-" sign is added to the VMM and hybrid class, indicating the pedestal problem (e.g. VMM Class "A" becomes "A-"). This requires to measure the pedestals at two different gains, once at \SI{3}{\milli\volt\per\femto\coulomb} and once at \textbf{\SI{16}{\milli\volt\per\femto\coulomb}}. Figure \ref{fig:pedestalProblem} shows the result of a pedestal measurement at the highest gain for a hybrid, on which one of the two VMM chips (VMM1, red curve) has the pedestal problem.
\subsection{Production yields}\label{sec:prod_yield}
\begin{figure}[htbp]
    \centering
    \includegraphics[width=0.49\linewidth]{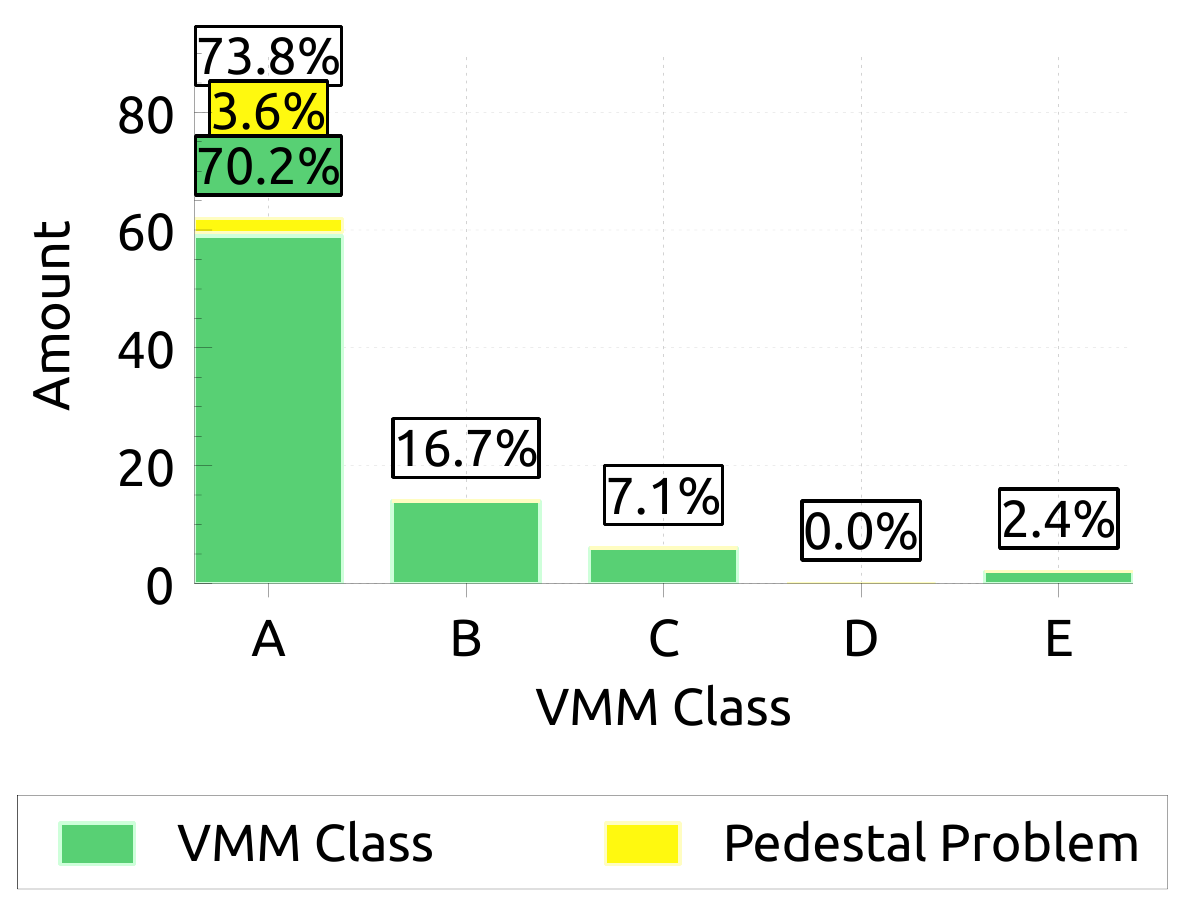}
    \includegraphics[width=0.49\linewidth]{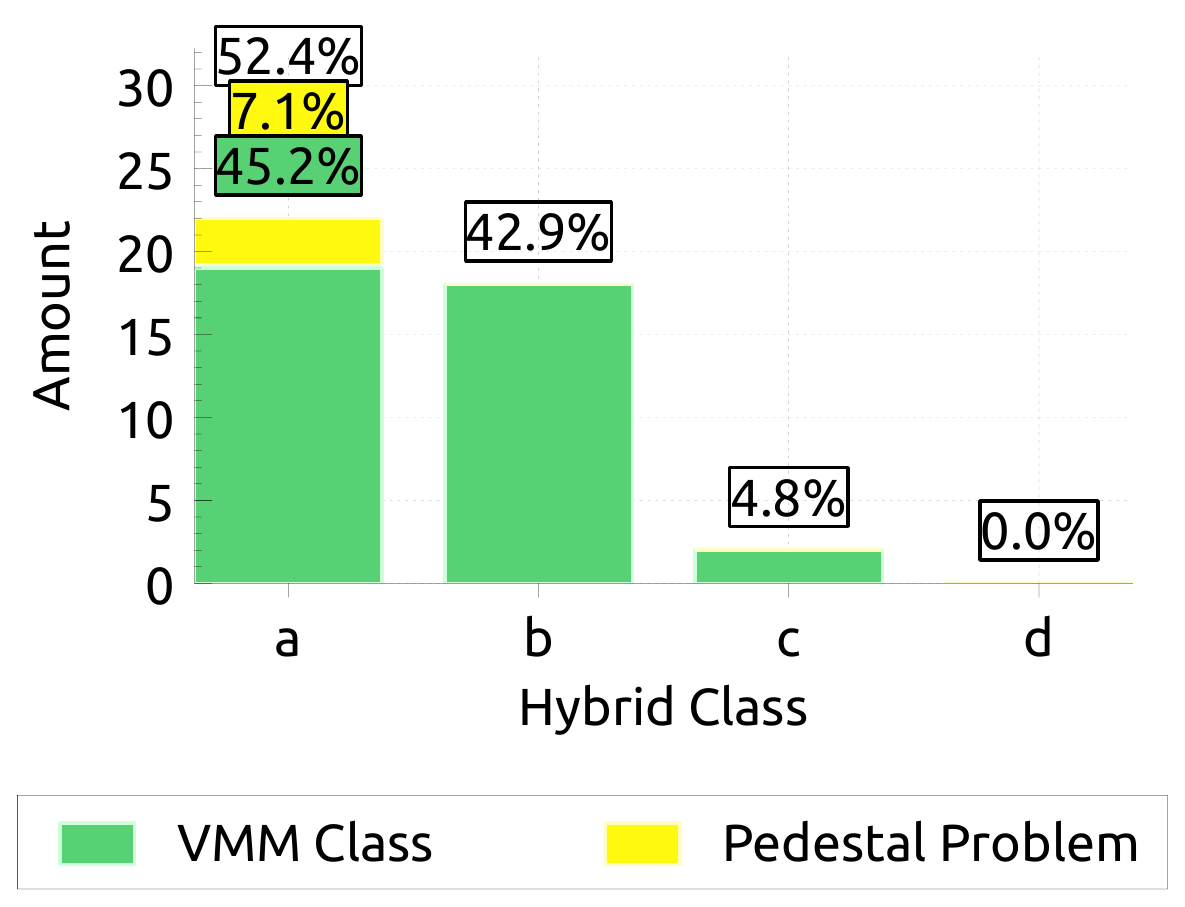}
    \caption{VMM and hybrid class distribution of 42 hybrids from production batch 1 (06/2020)}
    \label{fig:YieldSpring}
\end{figure}

\begin{figure}[htbp]
    \centering
    \includegraphics[width=0.49\linewidth]{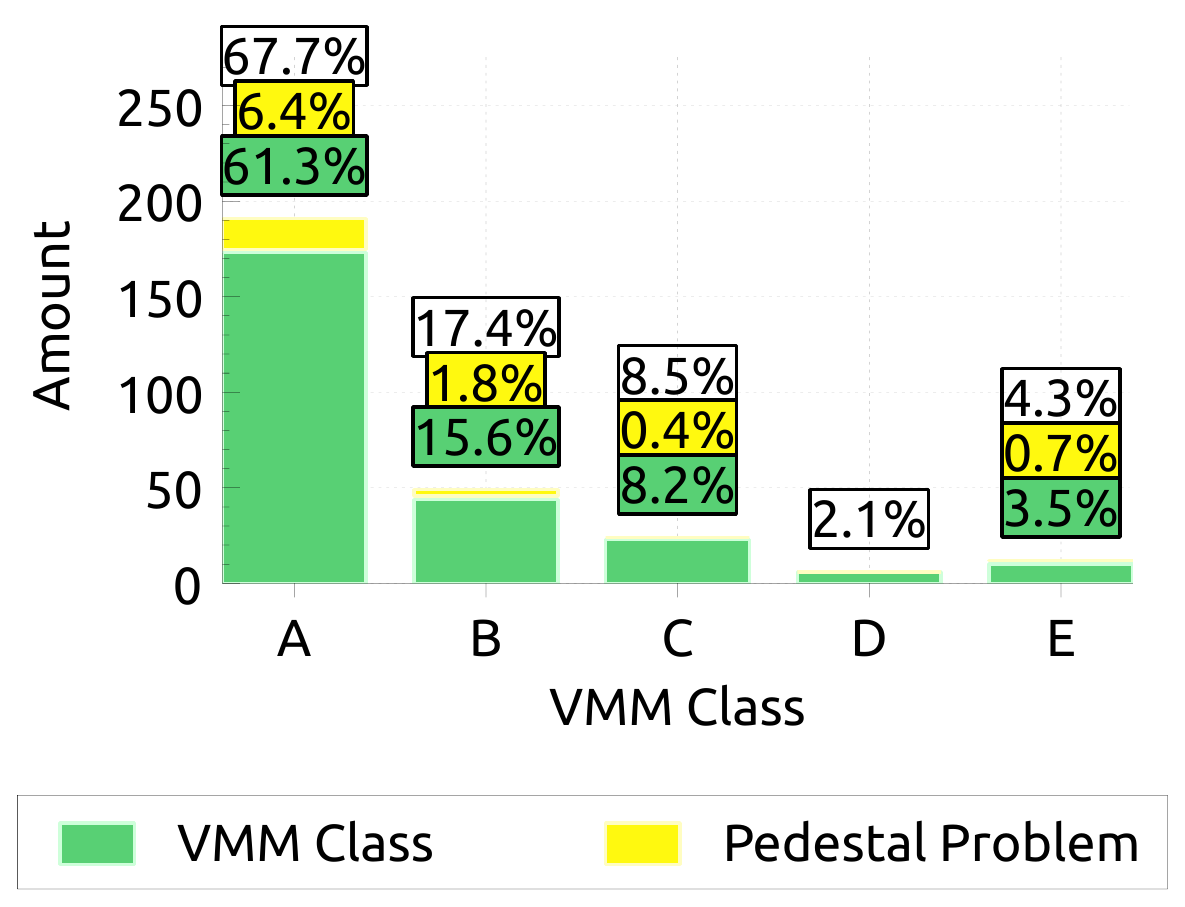}
    \includegraphics[width=0.49\linewidth]{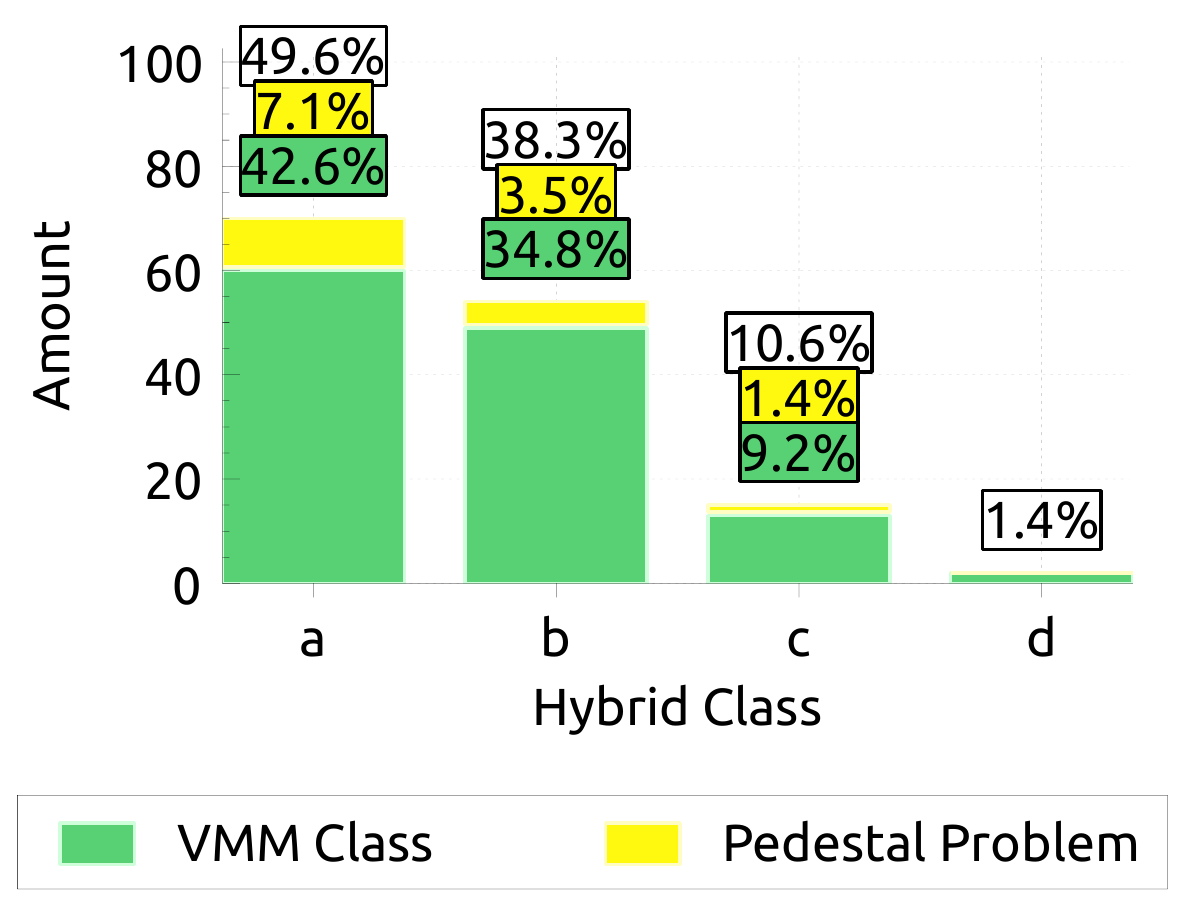}
    \caption{VMM and hybrid class distribution of 134 hybrids from production batch 2 (02/2021)}
    \label{fig:YieldAutumn}
\end{figure}

A total of 176 hybrids, of which 42 were produced in 06/2020 (batch 1) and 134 in 02/2021 (batch 2), were tested with the test system presented in this work. Using a database browser, the results of all tested hybrids can be evaluated. The tool offers a variety of filtering options, e.g. for the location the test was performed. For each hybrid, the full history of tests is accessible. The VMM and hybrid class distributions for both production batches are shown in figures \ref{fig:YieldSpring} and \ref{fig:YieldAutumn}. The numbers above the histogram bars indicate the percentage of VMMs or hybrids in the respective class including those with pedestal problems (white box) and excluding those with pedestal problems (green). The yellow bars and numbers indicate the percentage of hybrids with pedestal problems.
For the single VMM chip class A,B and C are considered to be acceptable (up to three not perfect channels per chip) and the sum of these classes make up the total yield. For the hybrids class a and b are acceptable (up to three not perfect channels per hybrid). With this the yields are determined to be for the 06/2020 production:
\begin{align*}
    Y_\text{VMM} &= \SI{97.6}{\percent} & Y_\text{Hybrid} = \SI{95.2}{\percent}
\end{align*}
and for the 02/2021 production
\begin{align*}
    Y_\text{VMM} &= \SI{93.6}{\percent} & Y_\text{Hybrid} = \SI{88.0}{\percent}
\end{align*}
If the pedestal problem would not be treated separately and all chips with these issues were rated as class E, the yields would drop e.g. for the batch 2 hybrids to \SI{77.4}{\percent}.
\section{Impact and outlook}\label{sec:Impact}
Comparing both production batches, some differences in the class distributions can be observed. However, due to the low number of chips in batch 1, they are statistically not significant assuming a random hybrid selection and $\sqrt{N}$ errors. A reason why the yields in batch 1 are better compared to those in batch 2 could be the lower number of hybrids and the fact that for the first production, every single hybrid was taken care of manually, which has to be automatised for larger productions.

An option to increase the yield would be to test all VMM chips already on the wafer and only use good chips for production, as was done for the ATLAS NSW project. However, this system only monitors the analogue behaviour of the chip and therefore would not detect all issues. The effort and cost to develop, maintain and operate such a system was elaborated by some developers in the RD51 Collaboration electronics group. Considering all these additional efforts, a limit for the hybrid yield of \SI{85}{\percent} was deduced, below which the strategy of wafer testing would have been beneficial. Based on the results of this study, wafer testing is not necessary.

With the automated testing system as described in this manuscript, several modes of failure for the VMM ASIC were found. Not all of those could be detected by the first VTC version during production. As a result, the VTC was further developed to detect e.g. problems with the VMM ADC, which primarily contributes to the E classification of chips without pedestal problems. It was decided to use the improved VTC during several steps of production to replace VMM chips with defects. With these improvements, a production yield of \SI{95}{\percent} for the hybrids should be achievable.


\section{Conclusion}\label{sec:Conc}
An automated testing system for the RD51 VMM hybrid was developed. It uses the standard SRS readout chain for the principle tests to qualify the hybrids. Optional extensions allow for additional tests e.g. on the power consumption or connectivity to a detector. The system is implemented in the VMM slow control and allows simple testing with graphical support. Test results are uploaded to a general database. Based on measurements on the first two VMM hybrid production batches, the respective yields have been measured. From the first production batch of 06/2020, \SI{95.2}{\percent} of the hybrid were rated acceptable (class a and b, see section~\ref{sec:prod_yield}). For the second production batch, \SI{88.0}{\percent} of the hybrids were rate acceptable. 
With these results, an ASIC testing on the wafer, for which a yield below \SI{85}{\percent} was considered worth the additional efforts by the RD51 Collaboration, was discarded.
Taking into account a pedestal problem only occurring for high VMM internal preamplifier gains, the yield would drop to \SI{77.4}{\percent} for the second production. 
A detailed analysis of the VMM failures lead to improved testing during the production with another testing system. It is expected that the yield of future productions will further increase.
\section{Acknowledgements}\label{sec:Ack}
We would like to thank SRS Technology and the RD51 community, especially the Working Group 5.1 members, for the fruitful discussions. 
We also recognise the assistance by Stefano Caiazza and Pepe G\"ulker from Mainz University and Ignacio L\'{a}zaro Roche from the LSBB for allowing us to test their hybrids to increase statistics. We would like to acknowledge the assistance by Dorothea Pfeiffer in answering questions on the VMM and developing human experience-based quality criteria.
Parts of this work have received funding from the European Union’s Horizon 2020 research and innovation programme under the Marie Sklodowska-Curie grant agreement no. 846674
as well as from the German Federal Ministry of Education and Research under grant no. 05K19PD1 and no. 13E18CHA (Wolfgang Gentner Programme).

\bibliography{references}

\end{document}